\documentclass[aps,amsfonts,prl,nofootinbib]{revtex4}

\usepackage{epsfig}

\input BoxedEPS
\SetRokickiEPSFSpecial  
\HideDisplacementBoxes

\newcommand{\half}{\fract{1}{2}}
\newcommand{\sech}{\mbox{sech}}

\newcommand{\fract}[2]{{\textstyle\frac{#1}{#2}}}

\begin{document}

\title{Finite Energy Sum Rules in Potential Scattering}
 
\author{N.~Graham,\footnote{
e-mail:  graham@physics.ucla.edu, jaffe@mit.edu,
maqua@mitlns.mit.edu,\\ herbert.weigel@uni-tuebingen.de}$^{\rm a}$
R.~L.~Jaffe,$^{\rm b}$
M.~Quandt,$^{\rm b}$ and
H.~Weigel\footnote{Heisenberg Fellow}$^{\rm c}$}

\affiliation{{~}\\$^{\rm a}$Department of Physics and Astronomy \\
University of California at Los Angeles, Los Angeles, CA  90095 \\
\\
$^{\rm b}$Center for Theoretical Physics, Laboratory for Nuclear 
Science and Department of Physics, \\
Massachusetts Institute of Technology \\
Cambridge, Massachusetts 02139 \\
\\
$^{\rm c}$Institute for Theoretical Physics~~ T\"ubingen University~~
Auf der Morgenstelle 14, D--72076 T\"ubingen, Germany\\
{~} \\
{\rm MIT-CTP\# 3128 \qquad UNITU-HEP-13/2001 \qquad quant-ph/0104136}
}

\begin{abstract}
We study scattering theory identities previously obtained as consistency
conditions in the context of one-loop quantum field theory
calculations.  We prove the identities using Jost function techniques and
study applications.
\end{abstract}

\maketitle

\section{Introduction}

In a recent study of the quantum energies of interfaces in field theory, we
discovered a set of consistency conditions on scattering data that take the
form of finite energy sum rules in potential scattering, \cite{Graham:2001dy}
\begin{equation}
	\int_0^\infty \frac{dk}{\pi} \, k^{2 n}
	\frac{d}{dk}\left[\delta_\ell(k)-\sum_{\nu=1}^m\delta_\ell^{(\nu)}(k)
	\right] + \sum_j (-\kappa_{\ell j}^{2})^{n} = 0\,, \qquad\quad m
	\ge n \,.
	\label{main}
\end{equation}
Here $\delta_\ell(k)$ denotes the scattering phase shift in the channel
with angular momentum $\ell$, and $\delta_\ell^{(\nu)}(k)$ is the $\nu^{\rm
th}$ Born approximation.  The sum on $j$ ranges over the bound states with
angular momentum $\ell$ and $\kappa_{\ell j}^{2}=-k_{\ell j}^{2}$ is the
binding energy.  Note that for $n=m=0$, eq.~(\ref{main}) is simply
Levinson's theorem.  In fact, these identities are the natural
generalizations of Levinson's theorem. $m=n$ is the 
minimal number of Born subtractions necessary to render the integral
in eq.~(\ref{main}) finite. Since we may generally subtract further
Born approximations, eq.~(\ref{main}) also implies ``oversubtraction" 
rules such as
\begin{equation}
	\int_0^\infty \frac{d k}{\pi}\,k^2\,\frac{d}{d k}\delta_\ell^{(2)}(k)
	 = 0\,. 
\end{equation}

In Ref.~\cite{Graham:2001dy} these sum rules appeared as consistency
conditions in quantum field theory  leading to finite expressions for 
the Casimir energies of interfaces.  Here we derive them within scattering
theory, and consider applications and consequences in ordinary quantum
mechanics.  Our proof will employ Jost function techniques commonly used to
prove Levinson's theorem \cite{scattering}.  For sufficiently singular
potentials, however, our sum rules fail, even though Levinson's theorem
continues to hold.

In Section~II we derive the sum rules for the antisymmetric channel in one
dimension, where the analysis is simplest.  This derivation applies to the
$s$-wave in three dimensions as well.  The extension to higher partial
waves in three dimensions is straightforward and is presented in the
Appendix.  In Section~III we discuss generalizations.  The generalization
to fermion scattering (via the Dirac equation) is also straightforward and
is left to the reader.  We also mention the generalization to multichannel
problems with internal symmetries.  The symmetric channel in one dimension
requires special consideration (as it does for Levinson's theorem
\cite{Barton:1985py}) and is treated in detail in Section~IV. In Section~V 
we describe some singular potentials for which the sum rules do not hold. 
Finally, in Section VI we study the semiclassical limit, where the sum
rules take a particularly compact form and have a simple physical
interpretation.

These sum rules could have been derived many years ago in the heyday of
potential scattering theory.  However, we have been unable to find them in
the literature.  They bear some resemblance to results based on the
Gel'fand-Diki\u{\i} equation, obtained in the Russian 
literature~\cite{Faddeev}, although the physical foundations and the 
resulting sum rules themselves are quite different. In our conclusion 
we mention this earlier work and compare it with our own. 
Also, for the special case of a separable potential a
related sum rule was obtained in Ref.~\cite{Jaffe:2000zp}.

\section{The Antisymmetric Channel in One Dimension}

\subsection{Derivation of the Sum Rules}

We consider the scattering of a spinless particle in a symmetric
 potential $V(x)=V(-x)$ in one dimension, described by the
Schr\"odinger equation,
\begin{equation}
	-\psi''+ V(x)\psi = k^2 \psi.
	\label{Schr1}
\end{equation}
This is a two channel problem.  The antisymmetric channel is specified by
the boundary condition $\psi_{-}(0)=0$.  The symmetric channel corresponds
to $\psi_{+}'(0)=0$.  Here we consider the antisymmetric channel.  Let
$\delta_{-}(k)$ denote the scattering phase shift in this channel, defined
by the asymptotic form of the  wave function $\psi(x)$ at large $x$,
\begin{equation}
	\psi_{-}(x)\to e^{-ikx}-e^{2i\delta_{-}(k)}e^{ikx}.
	\label{smatrix}
\end{equation}

Our goal is to derive the sum rules
\begin{equation}
	\int_0^\infty \frac{dk}{\pi} \, k^{2 n}
	\frac{d}{dk}\left[\delta_{-}(k)-\sum_{\nu=1}^m\delta_{-}^{(\nu)}(k)
	\right] = - \sum_j (-\kappa_{-,j}^{2})^{n} \,, \qquad\quad m
	\ge n
	\label{main1}
\end{equation}
where the sum on $j$ ranges over the antisymmetric bound states of
$V(x)$ with binding energies $\kappa_{-,j}^{2}=-k_{-,j}^{2}$.  For
the remainder of this section we suppress the subscript labeling
the antisymmetric channel.  For real $k$, the phase shift $\delta(k)$
is given in terms of the S-matrix $S(k)$, which in turn is related to
the Jost function
$F(k)$ by
\begin{equation}
	\delta(k) = \frac{1}{2 i}\ln S(k) = \frac{1}{2
	i}\left[\ln F(-k) - \ln F(k)\right] \,.
\end{equation}
The Born approximation is an expansion of the phase shift
$\delta$ (not the Jost function $F$ itself) in powers
of the interaction $V(x)$.  
The Jost function is obtained from the \emph{Jost solution} $f(k,x)$
to eq.~(\ref{Schr1}), which is asymptotic to an outgoing wave at
infinity,
\begin{equation}
	\lim_{x\to\infty}e^{-ikx} f(k,x) = 1,
\end{equation}
and $F(k)=f(k,0)$.  As is well known, the integral equation for $f(k,x)$
has a unique solution in the upper half $k$-plane, where it is holomorphic
and continuous as $\mathsf{Im}\,k \to 0$, provided that the potential
$V(x)$ is locally integrable and from the so-called ``Faddeev class"
\begin{equation}
	\int\limits_{-\infty}^\infty dx\,(1+|x|)\,|V(x)| < \infty\,.
	\label{faddeev}
\end{equation} 
In addition, $F(k)$ has zeros at the bound states, $k=i\kappa_{j}$, 
on the positive imaginary axis.\cite{scattering} To quantitatively
estimate the behavior of the Born approximation at large momenta $|k|$, 
we furthermore have to assume that the interaction $V(x)$ is bounded and 
sufficiently smooth to allow for integration by parts. Unless stated 
otherwise, we will restrict our analysis to non-singular potentials 
$V(x)$ with these properties.

To proceed, we take $m\geq n$ and introduce an auxiliary function,
$F_{m}(k)$, with the following properties:
\begin{enumerate}
	\item [(a)] $F_{m}(k)$ is analytic and has no zeros in the upper half
	$k$-plane including $k=0$.
	\item [(b)] $|\ln F(k)-\ln F_{m}(k)|$ falls like $|k|^{-2m-1}$ as 
	$|k|\to\infty$ in the upper half  plane. 
\end{enumerate}
After completing the derivation of our sum rules we will construct
$F_{m}(k)$ and relate it to the Born approximation to $\delta(k)$. 
For real $k$ we introduce
\begin{equation}
	\delta_{m}(k)\equiv \frac{1}{2
	i}\left[\ln F_{m}(-k) - \ln F_{m}(k)\right] \, ,
	\label{deltam}
\end{equation}
and consider
\begin{eqnarray}
	I_{n,m}&=&\int_{0}^{\infty}\frac{dk}{\pi}k^{2n}\frac{d}{dk}\left(
	\delta(k)-\delta_{m}(k)\right) \nonumber\\
        &=& - \frac{1}{2\pi i}\int_0^\infty dk\,k^{2n} \,\frac{d}{d k}
        \left(\ln F(k) - \ln F(-k) - \ln F_m(k) + \ln F_m(-k)\right)\ . \nonumber
\end{eqnarray}
Since the integrand is manifestly even in $k$, we can extend the integration
range to $-\infty$. Applying the substitution $k\to -k$ we obtain,
\begin{eqnarray}
	I_{n,m} &=&-\frac{1}{2\pi i}\int_{-\infty}^{\infty}dk \ k^{2n}\frac{d}{dk}
	\left(\ln F(k)-\ln F_{m}(k)\right)\cr
	&=&-\frac{1}{2\pi i}\oint_{\cal C}dk \ k^{2n}\frac{d}{dk}
	\left(\ln F(k)-\ln F_{m}(k)\right)\,.
	\label{contour}
\end{eqnarray}
where the contour ${\cal C}$ is the real axis plus the semicircle of infinite
radius in the upper half plane.  The semicircle gives no contribution
to the integral because of property (b).

The contour integral can now be performed using Cauchy's theorem by
recognizing that $d\ln F/dk$ has poles of unit residue at each bound
state.  By property (a), $d\ln F_m/dk$ is has no poles inside ${\cal C}$.  
The result is the sum rule, eq.~~(\ref{main1}).

\subsection{Construction of the Auxiliary Function}

In this section we construct an auxiliary function with the two
properties required in the previous subsection.  It is convenient to
parameterize the Jost solution $f(k,x)$ in terms of an exponent
$\beta(k,x)$,
\begin{equation}
	f(k,x) \equiv e^{i kx + i\beta(k,x)}\,. 
	\label{beta}
\end{equation}
Substituting into the Schr\"odinger equation we find that the
complex function $\beta(k,x)$ satisfies
\begin{equation}
	- i \beta''(k,x)+ 2 k \beta'(k,x) + \beta'^2(k,x) + V(x) = 0,
	\label{betaeq}
\end{equation}
subject to the boundary condition
\begin{equation}
	\beta(k,\infty) = \beta'(k,\infty) = 0\ ,
	\label{betabc}
\end{equation}
where $\beta'(k,x)= d\beta(k,x)/dx$.  Combining eqs.~(\ref{smatrix})
and (\ref{beta}) with the boundary condition $\psi(0)=0$, it is easy to
see that
\begin{equation}
	\delta(k)=-{\rm Re}\ \beta(k,0) \,.
	\label{deltabeta}
\end{equation}

Eqs.~(\ref{betaeq}) and (\ref{betabc}) can be converted into a
non-linear integro-differential equation,
\begin{equation}
	\beta(k,x) = \frac{1}{2k}\int_{x}^{\infty}dy\,\left( 
	1-e^{2ik(y-x)}\right)\Gamma(k,y),
	\label{betaint}
\end{equation}
where
\begin{equation}
    	\Gamma(k,x) = \beta'^{2}(k,x)+V(x).
	\label{gammadef}
\end{equation}
Note that  by differentiation $\beta'(k,x)$ obeys a similar equation,
\begin{equation}
	\beta'(k,x)=i\int_{x}^{\infty}dy\, e^{2ik(y-x)}\Gamma(k,y).
	\label{betaprime}
\end{equation}
Denote the term in $\beta(k,x)$ that is $\nu^{\rm th}$ order in the
potential by $\beta^{(\nu)}(k,x)$ and define
$\beta^{(\nu)}(k)\equiv\beta^{(\nu)}(k,0) $.  An equation for
$\beta^{(\nu)}(k,x)$ can be obtained from eq.~(\ref{betaint}) by
iteration
\begin{equation} 
	\beta^{(\nu)}(k,x) = \frac{1}{2k}\int_x^\infty dy
	\left(1-e^{2ik(y-x)}\right) \Gamma^{(\nu)}(k,y) \,.
	\label{Bornbeta}
\end{equation}
Here $\Gamma^{(\nu)}$ is the term in the expansion of $\Gamma$ which
is of $\nu^{\rm th}$ order in the potential. For $\nu>1$, 
$\Gamma^{(\nu)}$ involves only $\beta'^{(\mu)}$ with
$\mu<\nu$.  Thus we are led to equations for $\beta^{(\nu)}(k)$, the 
first few of which are
\begin{eqnarray}
	\beta^{(1)}(k) & = & \frac{1}{2k}\int_{0}^{\infty}dy\,\left( 
	1-e^{2iky}\right)V(y),\cr
	\beta^{(2)}(k) & = & \frac{1}{2k}\int_{0}^{\infty}dy\,\left( 
	1-e^{2iky}\right)[\beta'^{(1)}(k,y)]^{2},\cr
	\beta^{(3)}(k) & = & \frac{1}{2k}\int_{0}^{\infty}dy\,\left( 
	1-e^{2iky}\right)2\beta'^{(1)}(k,y)\beta'^{(2)}(k,y)\, ,
	\label{bornseries}
\end{eqnarray}
and so forth.  Similarly for the $\beta'^{(\nu)}(k,x)$, which appear as
sources in eqs.~(\ref{bornseries}),
\begin{eqnarray}
	\beta'^{(1)}(k,x) & = & i\int_{x}^{\infty}dy\,
	e^{2ik(y-x)}V(y),\cr
	\beta'^{(2)}(k,x) & = & i\int_{x}^{\infty}dy\,
	e^{2ik(y-x)}[\beta'^{(1)}(k,y)]^{2},\cr
	\beta'^{(3)}(k,x) & = & i\int_{x}^{\infty}dy\,
	e^{2ik(y-x)}2\beta'^{(1)}(k,y)\beta'^{(2)}(k,y)\, .
	\label{bornprimeseries}
\end{eqnarray}
The exponential factors in eq.~(\ref{Bornbeta}) guarantee that
$\beta^{(\nu)}(k)$ is analytic in the upper half $k$-plane provided that 
$\Gamma^{(\nu)}$ is, and likewise for $\beta'^{(\nu)}(k,x)$. Starting
with $\Gamma^{(1)}=V(x)$ we derive the required analytic properties
of $\beta^{(\nu)}$ and $\beta'^{(\nu)}$ inductively. To obtain the 
large $|k|$ behavior of $\beta^{(\nu)}$ and $\beta'^{(\nu)}$ from 
their respective integral representations, we integrate by parts 
once and estimate the remainder by sequentially applying the 
Riemann-Lebesgue lemma.\footnote{This procedure is allowed by our 
assumptions on the potential $V(x)$. In particular, the Riemann-Lebesgue
lemma requires the existence of $\langle |V|^\nu \rangle = \int_0^\infty
|V(y)|^\nu\,dy$ for all $\nu$, as well as similar averages involving
the derivative, e.g.~$\langle |V\,V'|\rangle < \infty$.} 
The result is that
\begin{itemize}
	\item For the class, eq.~(\ref{faddeev}) of potentials,  
        $\beta^{(\nu)}(k)$ is holomorphic in the upper half plane 
	including at $k=0$; 
	\item $|\beta^{(\nu)}(k)|\to \mathrm{const}\cdot |k|^{-2\nu+1}$ 
        as $|k|\to\infty$ 
        in the upper half plane $\mathsf{Im} \, k > 0$.
\end{itemize}
To complete the derivation, we define 
\begin{equation}
	F_{m}(k)= 
	\exp\left[i\sum_{\nu=1}^{m}\beta^{(\nu)}(k)\right].
	\label{aux}
\end{equation}
The required properties of $F_{m}(k)$ follow directly from those of 
$\beta^{(\nu)}(k)$ \and the convergence of the Born series 
$\beta(k) = \sum_{\nu=1}^\infty \beta^{(\nu)}(k)$ for sufficiently large 
$|k|$ in the upper half plane \cite{scattering,sabatier} 
(see also the Appendix).

The quantity that enters the sum rule is
$\delta_{m}(k)$, given by eq.~(\ref{deltam}).  From eq.~(\ref{betaeq}), it
follows that $\beta^{(\nu)}(-k)=-\beta^{(\nu)*}(k)$ for real $k$.  
As a result, we have
\begin{equation}
	\delta_{m}(k) = - {\rm Re}\ \sum_{\nu=1}^{m}\beta^{(\nu)}(k),
	\label{deltais}
\end{equation}
so that $\delta_{m}(k)$ is the sum of the first $m$ terms 
in the Born expansion of $\delta(k)$.  This completes the derivation 
of the sum rules in the antisymmetric channel in one dimension.

\section{Generalizations of the Basic Result}

The antisymmetric channel in one dimension is actually generic.  The 
sum rules  can easily be extended to scattering from a central potential
in any number of space dimensions $D>1$.  The computation proceeds 
for each partial wave in analogy to the antisymmetric case in $D=1$.  
Of course, the appropriate generalized Hankel functions must replace 
the simple exponentials that appear in one dimension.  We summarize 
the derivation for $D=3$ in the Appendix.

The sum rules also extend to the case of fermion scattering in a
straightforward way.  For a scalar potential, the Dirac equation
decomposes into partial waves labeled by total spin $j$ and parity
$\Pi$, and the sum rules again hold in each partial wave individually.

When there are internal symmetries, so that there are several
channels~$\{s\}$ in each partial wave, we expect that the sum rules will
continue to hold with the phase shifts replaced by the sum of the
eigenphases, which is given by the trace of the logarithm of S:
\begin{equation}
	\delta(k)\to\sum_{s}\delta_s(k)=\frac{1}{2i}{\rm Tr}\ln S(k) \,.
	\label{trln}
\end{equation}
Similarly, the sum over bound states will include all bound states in
the channel.  For example, consider an isodoublet of fermions
scattering in a background generated by an isodoublet scalar Higgs
field in three dimensions.  If the Higgs background is of the
``hedgehog'' form $\phi(\vec x)=\phi_0\,\exp(i\vec\tau\cdot
\vec{x}\,f(r))$, then the fermion spectrum will decompose into channels
labeled by parity $\Pi$ and grand spin $G$.  In each channel, $S$ is a
2-by-2 matrix, which cannot be simultaneously diagonalized for all $k$. 
(Each degree of freedom also appears with the usual $2G+1$ degeneracy.)  If
we introduce a chiral SU(2) gauge field that maintains grand spin
conservation, then states with different parity but the same $G$ will mix,
leaving a 4-by-4 $S$-matrix labeled only by $G$.

The symmetric channel in one dimension introduces additional subtleties,
which are treated in the following section.  The  result is that the sum
rules  may be modified by an anomalous piece if too many subtractions are
attempted.  Specifically, the sum rules in the symmetric channel read
\begin{equation}
	\int_0^\infty \frac{dk}{\pi} \, k^{2 n}
	\frac{d}{dk}\left[\delta_{+}(k)-\sum_{\nu=1}^m\delta_{+}^{(\nu)}(k)
	\right] = - \sum_j (-\kappa_{+,j}^{2})^{n} +I_{n,m}^{\rm anom}\,, 
	\label{mainsym}
\end{equation}
for $m\ge n$.  The anomalous term vanishes if
$2n>m$.  As a result, the ``minimally subtracted'' form of the sum rules,
where $m=n$, hold without modification except for the case
$m=n=0$, which is Levinson's theorem.  In that case $I_{0,0}^{\rm
anom}=\half$ and we recover the extra term that appears in Levinson's
theorem in the symmetric channel \cite{Barton:1985py}.

We have checked these results numerically in a variety of simple, generic
potentials. In one dimension, we have also checked them for the exactly
solvable reflectionless scalar potentials of the form $V(x) = -\ell(\ell+1)
\sech^2 x$, with $\ell$ integer \cite{MF}, and the corresponding
potentials in the fermionic case \cite{Graham:1999qq,Graham:1999pp}.

\section{The Symmetric Channel in One Dimension}

\subsection{Derivation of the Sum Rules}

As in the case of Levinson's theorem, the symmetric channel requires
special attention.  The regular solution to the Schr\"odinger equation
obeying the boundary conditions, $\psi'(0)=0$ and $\psi(0)=1$, can be
written in terms of the Jost solution $f(k,x)$,
\begin{equation}
	\psi(k,x)=\frac{1}{2ki}\left[G(k)f(-k,x)-G(-k)f(k,x)\right],
\end{equation}
where $G(k)=df(k,x)/dx|_{x=0}$.  Comparing to the $S$-matrix 
parameterization as $x\to\infty$,
\begin{equation}
	\psi(k,x)\to e^{-ikx}+e^{2i\delta_{+}(k)}e^{ikx},
\end{equation}
we see that
\begin{equation}
	\delta_+(k) = \frac{1}{2 i}\ln S(k) = \frac{1}{2
	i}\left[\ln (-G(-k)) - \ln G(k)\right] \,.
\end{equation}
The derivation proceeds exactly in analogy to the antisymmetric channel
except, as we shall see, it is not possible to find an auxiliary function
that is regular at $k=0$.  Instead we introduce an auxiliary function
$G_{m}(k)$ with the following properties:
\begin{enumerate}
	\item [(a)] $G_{m}(k)$ is analytic and has no zeros in the upper
	half $k$-plane excluding $k=0$.  
	\item [(b)] $|\ln G (k)-\ln G_{m}(k)|$ falls like $|k|^{-2m-1}$ as
	$|k|\to\infty$ in the upper half plane.
	\item[(c)] At $k=0$, $k^{2n} d\,\ln G_{m}(k)/dk$ has a pole with residue 
	$2I_{n,m}^{\rm anom}$.
	\item[(d)] $I_{n,m}^{\rm anom}=0$ for $2n>m$.
\end{enumerate}
Then, using the properties of $G$ and $G_{m}$ we find,
\begin{eqnarray}
	I_{n,m}&=&\int_{0}^{\infty}\frac{dk}{\pi}\,k^{2n}\frac{d}{dk}\left(
	\delta_+(k)-\delta_{+,m}(k)\right) \cr
        &=& - \int_0^\infty \frac{d k}{2\pi i}\,
         k^{2n}\frac{d}{d k}\left[\ln G(k) - \ln(-G(-k)) - \ln G_m(k) + 
         \ln(-G_m(-k))\right] \cr
	&=&-\frac{1}{2\pi i}\int_{-\infty}^{\infty}dk \ k^{2n}\frac{d}{dk}
	\left(\ln G(k)-\ln G_{m}(k)\right) \cr
	&=&-\frac{1}{2\pi i}\oint_{\cal C}dk \ k^{2n}\frac{d}{dk}
	\left(\ln G(k)-\ln G_{m}(k)\right)\cr
	&=&  \left\{\begin{array}{ll}
	- \sum_j (-\kappa_{j}^{2})^{n} & \qquad2n > m\ge n\\
	- \sum_j (-\kappa_{j}^{2})^{n} +I_{n,m}^{\rm anom}
	& \qquad m\ge 2n
	\end{array}\right.
	\label{contoursymm}
\end{eqnarray}
where
\begin{equation}
	\delta_{+,m}(k)= \frac{1}{2 i}\left[\ln (-G_{m}(-k)) - \ln
	G_{m}(k)\right] \,.
\end{equation}
The factor of $2$ difference between the residue of the pole at $k=0$
and the anomalous term in the sum rule arises because the integration
contour~${\cal C}$ passes through $k=0$ and therefore captures only
half the residue.  Given the restrictions on $m$ and $n$ in
eq.~(\ref{contoursymm}), it is clearly possible to derive a
non-anomalous sum rule in the symmetric channel by making the minimal
subtraction, $m=n$.  The only exception is Levinson's theorem,
$n=m=0$, which we discuss in detail below.  Otherwise anomalies arise
if one attempts to ``oversubtract'' for a given $n$.  We consider
specific examples after constructing the auxiliary function
$G_{m}(k)$.

\subsection{Construction of the Auxiliary Function}

As in the antisymmetric channel, the auxiliary function is obtained 
from the expansion of the Jost function in powers of the potential.  
The difference is that the relevant Jost function is $G(k)$ defined 
by $G(k)=df(k,x)/dx|_{x=0}$.  Using the exponential parameterization 
of  eq.~(\ref{beta}), we find
\begin{equation}
	G(k)=i(k+\beta'(k))e^{i\beta(k)} \,.
\end{equation}
where, $\beta'(k)=d\beta(k,x)/dx|_{x=0}$.
The prefactor $k+\beta'(k)$ gives the difference from the antisymmetric
channel.  Comparing with eq.~(\ref{aux}) we are led to the ansatz
\begin{equation}
	\ln G_{m}(k) = \left[\ln(k+\beta'(k))\right]_{m}+\ln F_{m}(k)
	\label{Gansatz}
\end{equation}
where the notation $[X]_{m}$ is an instruction to make the formal 
expansion of $X$ in powers of the potential and 
keep all terms up to $m^{\rm th}$ order.  For example,
\begin{eqnarray}
	\left[\ln(k+\beta'(k))\right]_{0}&=&\ln k\cr
	\left[\ln(k+\beta'(k))\right]_{1}&=&
        \ln k+\beta'^{(1)}(k)/k\cr
	\left[\ln(k+\beta'(k))\right]_{2}&=&
        \ln k+\beta'^{(1)}(k)/k
        +[\beta'^{(1)}(k)]^{2}/k^{2}+\beta'^{(2)}(k)/k \,.
\end{eqnarray}
This process is necessary to reproduce the asymptotic behavior
of $G(k)$ at large $|k|$ as required by condition (b) above.  The
cost is the introduction of poles up to $m^{\rm th}$ order at $k=0$.

It is straightforward to verify that $G_{m}(k)$ defined in
eq.~(\ref{Gansatz}) satisfies requirements (a) and (b) above. 
The argument is essentially the same as for the antisymmetric
channel.  It is clear from the definition of $G_{m}(k)$ that its
contribution to the integral along the real axis, proportional
to $\delta_{+,m}(k)$ is just the sum of the first $m$ terms in the Born
approximation to the phase shift in the symmetric channel.

It remains to characterize the singularity in $G_{m}(k)$ at $k=0$. 
The term in the contour integral in eq.~(\ref{contoursymm}) which is
potentially singular at $k=0$ is proportional to
\begin{eqnarray}
	k^{2n}\frac{d}{dk}[\ln(k+\beta'(k)]_{m}&=&
	k^{2n}\left[\frac{1+\frac{d\beta'(k)}{dk}}{k+\beta'(k)}\right]_{m}
	\cr
	&=& 
	k^{2n-1}\left[\left(1+\frac{d\beta'(k)}{dk}\right)\sum_{p=0}^{\infty}\left(
	\frac{-\beta'(k)}{k}\right)^{p}\right]_{m}
	\label{expand}
\end{eqnarray}
An anomalous contribution to the sum rule will result if the $1/k$
singularities in the expansion of $(1+\beta'(k))^{-1}$ to $m^{\rm th}$
order in the potential overcome the prefactor of $k^{2n-1}$.  Since
the functions $\beta'^{(\nu)}(k)$ are all analytic in the vicinity of
$k=0$, the most singular term in eq.~(\ref{expand}) comes from the
term $k^{2n-1}(-\beta'^{(1)}(k)/k)^{m}$, which is singular if $m\ge
2n$.  If $m=2n$ there is a simple pole at $k=0$ from this term.  If
$m>2n$ there are poles of higher order as well.  It is
straightforward (but increasingly tedious) to pull out the residue of
the simple pole, which determines the anomalous contribution to the sum
rule.  Once having identified the residue, the expression for the
anomalous contribution to $I_{n,m}$ is
\begin{equation}
	I_{n,m}^{\rm anom}= \frac{1}{2}{\rm Res}\ k^{2n-1}
	\left[\left(1+\frac{d\beta'(k)}{dk}\right)\sum_{p=0}^{\infty}\left(
	\frac{-\beta'(k)}{k}\right)^{p}\right]_{m}
	\label{residue}
\end{equation}

We illustrate this result with some important special cases:

\begin{itemize}
	\item $n=m=0$: Levinson's Theorem
	
	For $n=m=0$ we need the coefficient of $1/k$ in the term zeroth order 
	in the potential in $d\ln(k+\beta'(k))/dk$, which is unity.  Thus 
	$I_{0,0}^{\rm anom}=1/2$, and we obtain Levinson's theorem in the
	symmetric channel:\cite{Barton:1985py}
	\begin{equation}
		\int_{0}^{\infty}\frac{dk}{\pi}\frac{d}{dk}\delta_+(k)
		=\frac{1}{\pi}(\delta_+(\infty)-\delta_+(0) )= \frac{1}{2}-\sum_{j}1 
	\end{equation}

	\item $n=m>0$: Minimal subtraction
	
	For $n\ne 0$, the minimum Born subtraction we can make in order to
	render $I_{n,m}$ convergent is to take $m=n$.  The most singular
	term in the expansion of eq.~(\ref{expand}) through $m^{\rm th}$
	order is proportional to $(-\beta'_{1}(0))^{m}/k^{m+1}$.  Thus the
	integrand goes like $k^{2n-m-1}$ near $k=0$.  This has no pole
	when $n=m>0$.  So $I_{n,n}^{\rm anom}=0$ for $n>0$, and the
	minimally subtracted form of the sum rules is not altered in the
	symmetric channel.
	
	\item $2n>m$: Oversubtraction without an anomaly:
	
	There is no singularity at $k=0$ as long as
	$2n>m$.  Therefore it is possible to subtract further Born
	approximations from the phase shift without introducing anomalies
	into the sum rules.  For $n=1$ the first Born approximation must
	be subtracted for convergence and no further subtraction is
	possible without anomaly.  For $n=2$ the first and second Born
	approximations must be subtracted and the third may be subtracted
	without anomaly, and so forth.
	
	\item $m=2n$: Computation of the anomaly.
	
	For fixed $n$, as further subtractions are attempted, one finally 
	reaches $m=2n$, where an anomaly appears.  The anomaly comes entirely 
	from the  term proportional to $(-\beta'^{(1)}(0))^{2n}$ in the expansion 
	of the integrand.  Referring back to the definition of 
	$\beta'^{(\nu)}(k)$, we find 
	\begin{equation}
		\beta'^{(1)}(0)=i\int_{0}^{\infty}dyV(y)
	\end{equation}
	So
	\begin{equation}
	I_{n,2n}^{\rm anom}=
        \frac{(-)^n}{2}\left[\int_{0}^{\infty}dyV(y)\right]^{2n}
	\end{equation}
	
	In particular, the first non-trivial case is the $n=1$ sum rule from
	where both the first and second terms in the Born approximation have
	been subtracted:
	\begin{equation}
	I_{1,2}=
	\int_0^\infty \frac{dk}{\pi} \, k^{2}
	\frac{d}{dk}\left[\delta_+(k)-\delta_+^{(1)}(k)-\delta_+^{(2)}(k) \right]
	= \sum_j \kappa_j^{2}
	-\frac{1}{2}\left[\int_{0}^{\infty}dyV(y)\right]^{2}
	\end{equation}
	This result was first discovered in conjunction with the work of
	\cite{Graham:2001dy} by direct evaluation of the Feynman graph
	corresponding to the second Born approximation.  Here we see 
	that it follows from a careful analysis of the analytic properties 
	of the Born approximation near $k=0$ and has essentially the same 
	origin as the extra factor of $1/2$ that appears in Levinson's 
	theorem for the symmetric channel.
\end{itemize}

\section{Singular Potentials}

Among the cases that we have checked numerically is the square well
in one dimension.  Even though its sharp edges seem to invalidate our use of
complex analysis in the proof above, the sum rules still hold.  As is the
case with Levinson's theorem, the difference between a sharp edge and
a smooth, very steep edge can be made arbitrarily small.  

If we take the limit where the width of the well goes to zero with the area held
fixed, we obtain the delta-function potential
\begin{equation}
	V(x) = -\lambda \delta_D(x)
\end{equation} 
where we have written the Dirac delta function as $\delta_D(x)$ to
distinguish it from a phase shift.  The phase shift in this potential
vanishes in the antisymmetric channel because $V(x)$ is localized at
the origin, where the antisymmetric wavefunction vanishes.  The
symmetric channel phase shift and its first Born approximation are
easily calculated:
\begin{eqnarray}
	\delta_+(k) &=& \arctan \frac{\lambda}{2 k}\nonumber\\
	\delta^{(1)}_+(k) &=& \frac{\lambda}{2 k}
\end{eqnarray}
The symmetric channel has a bound state at $\kappa =
\frac{\lambda}{2}$.  

Like the square well, this potential obeys the
one-dimensional version of Levinson's theorem \cite{Barton:1985py} relating
the phase shifts to the number of bound states in each channel
\begin{eqnarray}
	\delta_-(0) &=& \pi n_- = 0 \nonumber\\
	\delta_+(0) &=& \pi (n_+-\frac{1}{2}) = \frac{\pi}{2}.
\end{eqnarray}
However, sum rule with $m=n=1$, eq.~(\ref{mainsym}), fails.  One
expects $I_{1,1}=\kappa^{2}=\lambda^{2}/4$, but obtains instead
$I_{1,1}=\lambda^{2}/8$.  Examining a sequence of square well
potentials approaching the delta function reveals that for any square
well, the sum rule is satisfied, but the support of the integral moves
out to larger and larger $k$ as the potential gets narrower and
deeper.  The $\delta_{D}$-function limit and the $k$-integration do
not commute.

It is instructive to examine more closely what goes wrong in the
$\delta_{D}$-function case.  A straightforward calculation
shows that the proper Jost function for the symmetric channel in one
dimension is
 \begin{equation}
	 G(k)= ik+\lambda/2
 \end{equation}
Note that it has a zero at $k=i\lambda/2$ as expected and is analytic
in the upper half $k$-plane.  According to the symmetric channel
analysis,
\begin{eqnarray}
	\ln G_{0}(k) &=& \ln ik\nonumber\\
	\ln G_{1}(k) &=& \ln ik -i\lambda/2k
\end{eqnarray}
 The derivation of Levinson's theorem using $\ln G-\ln G_{0}$ proceeds
 without difficulty.  To derive the sum rule for $I_{1,1}$ it is
 necessary to consider $d(\ln G(k) - \ln G_{1}(k))/dk$.  This quantity
 vanishes like $1/k^{3}$ for large $|k|$, {\it not} $1/k^{4}$ as
 expected on the basis of property (b) listed in the previous section. 
 As a result, the integral around the semicircle at infinity does not 
 vanish.  Specifically,
 \begin{equation}
	 I_{1,1}^{\infty}= -\frac{1}{2\pi i}\int_{\cal C_{\infty}}dk
	 k^{2}\frac{d}{dk}\left[ \ln(ik+\lambda/2) - (\ln ik
	 -i\lambda/2k)\right] =\frac{\lambda^{2}}{8}
\end{equation}
where ${\cal C_{\infty}}$ is the semicircle at infinity in the upper
half $k$-plane.  We combine this result with the integral along the
real axis,
\begin{equation}
	I_{1,1}= -\frac{1}{2\pi i}\oint_{\cal C}dk k^{2}\frac{d}{dk}\left[
	\ln(ik+\lambda/2) - (\ln ik -i\lambda/2k)\right]-\lambda^{2}/8
\end{equation}
Now $I_{1,1}$ can be evaluated by contour integration, yielding the
same anomalous result obtained by direct integration of
$\delta-\delta^{(1)}$ along the real axis:
$I_{1,1}=\frac{1}{2}\kappa^{2} = \lambda^{2}/8$.
 
It remains to explain why $\frac{d}{dk}(\ln G(k) - \ln 
G_{1}(k))$ falls only like $1/k^{3}$.  Consider
\begin{equation}
	\ln G_{1}(k)=\ln k +\beta'^{(1)}(k)/k +\ln F_{1}(k).
\end{equation}
Since $\ln F_{1}(k)$ is proportional to $\beta^{(1)}(k)$,
$\beta^{(1)}$ and $\beta'^{(1)}$ determine the large $k$ behavior of
$G_{1}(k)$.  From their definitions, eqs.~(\ref{betaint}) and (\ref{betaprime}),
\begin{eqnarray}
 \beta^{(1)}(k) & = & \frac{1}{2k}\int_{0}^{\infty}dy\,\left( 
1-e^{2iky}\right)V(y) = 0,\nonumber\\
\beta'^{(1)}(k) & = & i\int_{0}^{\infty}dy\,
e^{2iky}V(y) = -i\lambda/2,
\end{eqnarray}
for $V(x)=-\lambda V_{D}(x)$.
In particular, as $k\to\infty$, $\beta'^{(1)}(k)\to{\rm const}$.  In
contrast, if $V(x)$ is any bounded function of $x$, including a square
well, $\beta'^{(1)}(k)\sim V(0)/k$.  This is the
ultimate source of the breakdown of the sum rule in the case of the
$\delta_{D}$ function, for which $V(0)$ is ill-defined.

\section{WKB Applications}

Our sum rules become especially simple 
in the WKB approximation.  They provide formulas for the sum of powers of
the binding energies as integrals over powers of the potential.  In this
way, the WKB approximation yields some insight into the physical origin of
the sum rules.  We have checked the accuracy of these results in some
simple potentials.

We work in one dimension with a potential $V(x)$ that is everywhere
negative.  We therefore define $U(x)=-V(x)$.  We assume that $\int dx
[U(x)]^{n}$ exists for all $n\geq\half$.  The reflection coefficient is
exponentially small in the WKB approximation, so the even and odd parity
phase shifts are equal and are given by
\begin{equation}
\delta(k)=\int_{0}^{\infty}dx \left(\sqrt{k^{2}+U(x)}-k\right).
\label{WKBvalid}
\end{equation}
The $\nu^{\rm th}$  Born approximation to $\delta(k)$ is merely the term of 
order $U^{\nu}$ in the expansion of the integrand.  

The WKB approximation should be valid when $d\lambda(x)/dx \ll 1$, where 
$\lambda(x)=1/\sqrt{k^{2}+U(x)}$ is the local de~Broglie wavelength.  
For a deep, smooth potential this criterion is satisfied for all $x$.
The first correction to the WKB approximation gives only a modulation of 
the magnitude of the wavefunction and does not change its phase.  So 
we expect eq.~(\ref{WKBvalid}) to be quite a good approximation.

\begin{figure}[hbt]
\centerline{\BoxedEPSF{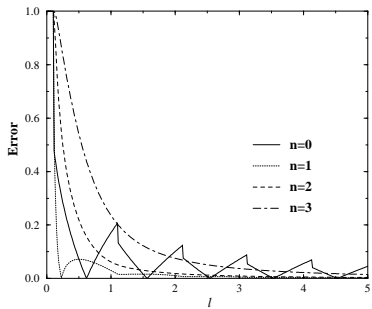 scaled 300}}\bigskip
\caption{\sl The relative error, defined as the difference of the 
right and left sides of eq.~(\ref{WKBresult}) divided by the sum,
for the potential $V(y) = -l(l+1)^2 \mathop{\rm sech}^2(y)$ 
as a function of the coupling $l$ for various sum rules that
are labeled by $n$.}
\label{wkbplot}
\end{figure}

To evaluate the sum rule with minimal subtraction ($m=n$) we must
calculate
\begin{equation}
I_{n,n} = \sum_{j}\kappa_{j}^{2n} =
(-1)^{n+1}\frac{2n}{\pi}\int_{0}^{\infty}k^{2n-1}
\left(\delta(k)-\sum_{\nu=1}^{n}\delta^{(\nu)}(k)\right) dk \,.
\end{equation}
A straightforward calculation yields the WKB estimate
\begin{equation} 
\sum_{j}\kappa_{j}^{2n} \approx 
\frac{2^{n+1}}{\pi}\frac{n!}{(2n+1)!!}
\int_{0}^{\infty}dy\left[U(y)\right]^{n+\frac{1}{2}} \equiv
I_{n,n}^{\rm WKB}\,.
\label{WKBresult}
\end{equation}
Note that in the WKB approximation, the $\nu^{\rm th}$ Born 
approximation is proportional to $1/k^{2\nu+1}$ so oversubtraction of 
the sum rules is not allowed in this case.
In Fig.~\ref{wkbplot} we show the relative error that arises
due to the WKB approximation for various sum rules. For sufficiently
strong potentials this error is indeed small.

Alternatively, we can use zeta function regularization to define a 
regularized integral 
\begin{equation} 
	I_{n,n}^{\rm WKB}(s) =
	(-1)^{n+1}\frac{2n}{\pi}\int_{0}^{\infty}k^{2n-1}\int_{0}^{\infty}dy
	\left(k^{2}+U(y)\right)^{s} dk
\end{equation}
and evaluate for $s<-n-1$, where it converges to a beta
function.  We then analytically continue to $s=\half$, and obtain the same
result.  Curiously, this result implies that in the WKB approximation, the
contribution of the Born terms vanish in zeta function regularization.

Note that $I_{0,0}^{\rm WKB}$ gives the WKB approximation to Levinson's theorem:
\begin{equation}
\sum_{j}  1 \stackrel{\rm WKB}{\approx} \frac{2}{\pi}\int_{0}^{\infty}dy\sqrt{U(y)},
\end{equation}
and $I_{1,1}^{\rm WKB}$ gives a particularly simple formula for the sum of the 
binding energy of all bound states in the WKB approximation:
\begin{equation}
\sum_{j}  \kappa_{j}^{2} \stackrel{\rm WKB}{\approx} 
\frac{4}{3\pi}\int_{0}^{\infty}dy\left[U(y)\right]^{\frac{3}{2}},
\end{equation}

Not surprisingly, eq.~(\ref{WKBresult}) has a simple semiclassical
interpretation.\footnote{We thank J.~Goldstone for this observation.}  We
replace the sum over bound states by an integral over the density of
states, $\rho(k)$,
\begin{equation}
I_{n,n}= \int d\rho(k) (-k^2)^{n}
\end{equation}
where the integral extends only over bound states ($k^{2}<0$).  If we
approximate $d\rho$ semiclassically by $d\rho=dydp/2\pi$, where
$k^{2}=p^{2}+V(y)$, then
\begin{equation}
I_{n,n}= 
\frac{1}{2\pi}\int_{-\infty}^{\infty}dy
\int_{-\sqrt{U(y)}}^{\sqrt{U(y)}}dp
\left(U(y)-p^{2}\right)^{n}\, .
\end{equation}
Direct evaluation of this simple integral yields eq.~(\ref{WKBresult}).

\section{Discussion and Conclusions}

Our sum rules are related to the results of Buslaev and Faddeev
based on the Gel'fand-Diki\u{\i} equation
\cite{Faddeev}.\footnote{We thank G.~Dunne for bringing this work to
our attention.} They studied solutions to the Schr\"odinger equation
on the half-line subject to the boundary condition $\psi(0) = 0$.  In
our language, this system is equivalent to the antisymmetric channel
in one dimension, though without the restriction that the potential be
smooth at the origin (so they can have $V'(0) \neq 0$).  They obtained
a sequence of sum rules, beginning with\footnote{We have corrected 
some apparently typographical sign errors in their results.}
\begin{eqnarray}
	\frac{2}{\pi} \int_0^\infty k \left(\delta(k) + \frac{1}{2k}\int
	V(x) dx \right) dk + \sum_j \kappa_j^2 &=& \frac{1}{4}V(0) \cr
	\frac{4}{\pi} \int_0^\infty k^3 \left(\delta(k) + \frac{1}{2k}\int
	V(x) dx +\frac{1}{(2k)^3}\left(V'(0) + \int V(x)^2
	dx\right)\right)dk - \sum_j \kappa_j^4 &=& \frac{1}{8}\left(2
	V(0)^2-V''(0)\right) \,.
\end{eqnarray}
Their identities have a similar structure to our sum rules (after
integrating by parts).  There are some significant differences, however.
Instead of subtracting the Born approximation as we have done, they instead
have subtracted the leading local asymptotic expansion of the phase shift
in powers of $1/k$.  Their expressions are simpler --- just integrals
of the potential over space divided by powers of $k$ --- but more
singular at the origin.  As a result, it is not possible to form the
oversubtracted versions of their identities.  This difference also accounts
for the need for extra terms proportional to the potential and its
derivatives at the origin.  In field theory applications, the 
Born subtractions arise naturally in the process of renormalization in 
a definite scheme.\cite{Graham:2001dy} We know of no similar application of the 
Buslaev-Faddeev results.

\section{Acknowledgements}
We thank E.~Abers, G.~Dunne, and J.~Goldstone for helpful discussions,
suggestions, and references.  M.~Q. and H.~W. are supported in part by
Deutsche Forschungsgemeinschaft under contracts Qu 137/1-1 and We
1254/3-1.  R.~L.~J. is supported in part by the U.S.~Department of Energy
(D.O.E.) under cooperative research agreement
\#DF-FC02-94ER40818.  N.~G. is supported by the U.S.~Department of Energy
(D.O.E.) under cooperative research agreement \#DE-FG03-91ER40662.

\section{Appendix}
\label{app}
In this section, we extend the proof of the sum rules to spherically
symmetric potentials in three dimensions.  Our analysis follows the case of
the antisymmetric channel in one dimension closely.

\subsection{Derivation of the Sum Rules}
We consider the scattering of a spinless particle in a central
potential $V(r)$ in three dimensions described by the radial 
Schr\"odinger equation,
\begin{equation}
	-\psi_\ell'' + \left[V(r) + \frac{\ell(\ell+1)}{r^2}\right]\,\psi_\ell =
	k^2	\psi_\ell \,.
	\label{Schr}
\end{equation}
Let $\delta_\ell(k)$ denote the scattering phase shift in the channel
with angular momentum $\ell$, and $\delta_\ell^{(\nu)}(k)$ the
$\nu^{\rm th}$ Born approximation.  Our goal is to derive the sum
rules
\begin{equation}
	\int_0^\infty \frac{dk}{\pi} \, k^{2 n}
	\frac{d}{dk}\left[\delta_\ell(k)-\sum_{\nu=1}^m\delta_\ell^{(\nu)}(k)
	\right] = - \sum_j (-\kappa_{\ell j}^{2})^{n} \,, \qquad\quad m
	\ge n
	\label{mainell}
\end{equation}
where the sum on $j$ ranges over the bound states with angular
momentum $\ell$ and $\kappa_{\ell j}^{2}=-k_{\ell j}^{2}$ is the
binding energy.  For real $k$, the phase shift $\delta_\ell(k)$ is
given in terms of the S-matrix $S_\ell(k)$, which in turn is related
to the Jost function
$F_\ell(k)$ by%
\begin{equation}
	\delta_\ell(k) = \frac{1}{2 i}\ln S_\ell(k) = \frac{1}{2
	i}\left[\ln F_\ell(-k) - \ln F_\ell(k)\right] \,.
\end{equation}
The Born approximation is an expansion of the phase shift
$\delta_\ell$ in powers of the interaction $V(r)$.

The Jost function is obtained from the Jost solution $f_\ell(k,r)$, which
is the solution to eq.~(\ref{Schr}) asymptotic to an outgoing wave at
infinity
\begin{equation}
	\lim_{r\to\infty}e^{-ik r} f_\ell(k,r) = i^{\ell}
	\label{jostbc}
\end{equation}
Uniqueness of $f_\ell(k,r)$ and analyticity in the upper 
$k$-plane can be verified for locally integrable potentials
$V(r)$ with the property \cite{scattering}
\begin{equation}
\int_0^\infty dr\,(1+r)\,|V(r)| < \infty.
\end{equation}
It is convenient to introduce the free Jost solution
\begin{equation}
	w_{\ell}(kr)=ikrh_{\ell}^{(1)}(kr)\,,
	\label{hankel}
\end{equation}
where $h_{\ell}^{(1)}(z)$ is the spherical Hankel function.  
$F_{\ell}(k)$ is determined by the $r\to 0$ limit of $f_{\ell}(k,r)$,
\begin{equation} 
	F_\ell(k) =\lim_{r\to0}\frac{f_\ell(k,r)}{w_\ell(kr)}\, .
	\label{jostdef}
\end{equation}
As in previous cases, $F_{\ell}(k)$ is analytic in the upper half 
$k$-plane with zeros at the bound state momenta, $k=i\kappa_{j}$.

The sum rules are derived once again by introducing an auxiliary 
function, $F_{\ell,m}(k)$, with the properties
\begin{itemize}
	\item [(a)] $F_{\ell,m}$ is analytic and has no zeros in the upper half 
	$k$-plane including $k=0$.
	\item [(b)] $|\ln F_{\ell}(k)-\ln F_{\ell,m}(k)|$ falls asymptotically 
	like $|k|^{-2m-1}$ as $|k|\to \infty$ in the upper half plane.
\end{itemize}
For real $k$ we define
\begin{equation}
	\delta_{\ell,m}(k)=\frac{1}{2i}\left[\ln F_{\ell,m}(-k)-\ln 
	F_{\ell,m}(k)\right]
	\label{deltalm}
\end{equation}
and the remainder of the derivation follows exactly as in the 
antisymmetric channel in one dimension:  The sum rule 
eq.~(\ref{mainell}) follows from Cauchy's theorem.

\subsection{Construction of the Auxiliary Function}

As in one dimension, it is convenient to parameterize
 $f_{\ell}(kr)$ in terms of an exponential,
\begin{equation}
	f_\ell(k,r) \equiv e^{i \beta_\ell(k,r)} w_{\ell}(kr)
	\label{betaell}
\end{equation}
Substituting into the radial Schr\"odinger equation, we find that the
complex function $\beta_\ell(k,r)$ satisfies
\begin{equation}
	- i \beta_\ell''(k,r) + 2 k \eta_\ell(k r) \beta'_{\ell}(k,r) +
	\beta_{\ell}(k,r)'^2 + V(r) = 0,
	\label{betaeqell}
\end{equation}
subject to the boundary condition
\begin{equation}
	\beta_{\ell}(k,\infty) = \beta_{\ell}'(k,\infty) = 0
	\label{betabcell}
\end{equation}
where $\beta_{\ell}'(k,r) = d\beta_{\ell}(k,r)/dr$. Here we have
introduced $\eta_{\ell}(z)$, 
\begin{equation}
	\eta_\ell(z) \equiv - i \,\frac{w_\ell'(z)}{w_\ell(z)}
	=- i \frac{d}{d z} \ln\left[z h^{(1)}_\ell(z)\right]\,,
\label{defetaell}
\end{equation}
which is a simple rational function of $z$.

It is convenient to convert eq.~(\ref{betaeqell}) into a non-linear
integro-differential equation,
\begin{equation} 
	\beta_\ell(k,r) = i \int_r^\infty dr_1\int_{r_1}^\infty dr_2\,
	\left(\frac{w_\ell(kr_2)}{w_\ell(kr_1)}\right)^{2}
	\Gamma_\ell(k,r_2)
        \label{betaint1}
\end{equation}
where
\begin{equation}
	\Gamma_{\ell}(k,r) = \left[V(r) + \beta_\ell'^2(k,r)\right] \,.
	\label{betaintell}
\end{equation}
By definition the value of $i\beta_\ell(k,r)$ at $r=0$ is the
logarithm of the Jost function,
\begin{equation} 
	F_\ell(k) =\lim_{r\to0}\frac{f_\ell(k,r)}{w_\ell(kr)}=
	e^{i \beta_\ell(k,0)}\, .
\end{equation}
Furthermore, the analytic properties of the Jost solution imply
$\beta(-k,r) = - \beta^\ast(k^\ast,r)$, yielding the phase shift
\begin{equation}
	\delta_\ell(k) = - \mathsf{Re}\,\beta_\ell(k,0)
	= \frac{1}{2}\left[\beta_\ell(-k,0) - 
	\beta_\ell(k,0)\right]
	\label{deltaeqell}
\end{equation}
for real $k$.
The Born series for $\delta_{\ell}(k)$ is constructed by iterating 
eq.~(\ref{betaeqell}) and keeping track of powers of $V(r)$ in the source
$\Gamma_\ell(k,r)$ on the r.h.s.~of (\ref{betaint1}).  With these definitions,
we are prepared to construct $F_{\ell,m}(k)$.  We begin by proving two
important properties of the $\{\beta^{(\nu)}\}$ using induction.

First, we need to show that the $\mathcal{O}(V^\nu)$
approximation $\beta_\ell^{(\nu)}(k, 0)$ of eqs.~(\ref{betaeqell}) and
(\ref{betabcell}) is a holomorphic function of $k$ in the upper half plane
$\mathsf{Im}(k) \ge 0$.  To see this, we go back to the initial 
value problem for $\beta_\ell$ and  rewrite eq.~(\ref{betaeqell}) as
\begin{equation}
	- i \beta_\ell''^{(\nu)}(k,r) + 2 k \,\eta_\ell(k r)
	\beta_\ell'^{(\nu)}(k,r) = - \Gamma_\ell^{(\nu)}(k,r)
	\label{betaelldiffeq}
\end{equation}
with the boundary condition
$\beta_\ell^{(\nu)}(k,\infty)=\beta_{\ell}'^{(\nu)}(k,\infty) = 0$
for all $k$.  $\Gamma_\ell^{(\nu)}$ is the $\mathcal{O}(V^\nu)$
term in the iteration of $\Gamma_\ell = V + \beta_\ell'^2$, including
all combinations of $V$ and $\beta_\ell'^{(\mu)}$ that give a total
order of $V^\nu$:
\begin{equation}
	\Gamma_\ell^{(1)}(k,r) = V(r)
\end{equation}
and
\begin{equation}
	\Gamma_\ell^{(\nu)}(k,r) = \sum_{\sigma+\tau = \nu}
	\beta_\ell'^{(\sigma)}(k,r) \beta_\ell'^{(\tau)}(k,r) 
	\qquad {\rm for }\, \nu\ge2\,.
	\label{Gammaexp}
\end{equation}
We proceed by induction and therefore assume that the right hand side of
eq.~(\ref{betaelldiffeq}) is holomorphic in the upper half $k$-plane.  This
is certainly the case for $\nu=1$.  Since the boundary condition is
independent of $k$, Poincar\'{e}'s theorem ensures that the solution
$\beta_\ell^{(\nu)}(k,r)$ of eq.~(\ref{betaelldiffeq}) is in fact a holomorphic
function of $k$ in every domain where the coefficients are.  It thus
remains to show that $\eta_\ell(z)$ as defined in eq.~(\ref{defetaell}) is
holomorphic in $z$ for $\mathsf{Im}\,z \ge 0$.  This will be the case if
the free Jost solution $w_\ell(z)$ is non-vanishing in the upper half
plane.  For $\ell=0$ it is trivial, since $w_0(z) = e^{i z}$.  For 
$\ell >0$, it suffices to note that any zero of $w_\ell$ in the upper half 
plane would correspond to a bound state of the free Schr\"odinger equation, 
which is forbidden by the repulsive centrifugal barrier.

Second, we have to establish the convergence of the
Born series. The iteration of the integral equation for
$f_\ell(k,r)$ yields an expansion in the interaction
$\sum_{\nu=1}^\infty f_\ell^{(\nu)}(k,r)$ that is uniformly and
absolutely convergent
in the upper half plane $\mathsf{Im}\,k > 0$. From the bound \cite{sabatier}
\[
|f_\ell(k,r) - w_\ell(k\,r)| < \left(\frac{|k| r}{1 + |k| r}\right)^\ell\,
\frac{e^{- \mathsf{Im}\, kr}}{|k|}\int_r^\infty |V(r_1)|\,d r_1
\]
and the limit $\lim_{|k|\to\infty} w_\ell(k\,r) = 1$
in the upper half plane $\mathsf{Im}\,k \ge 0$, it is clear that for
any given $r \ge 0$, we can always find a radius $\rho_r$ such that
$|F_\ell(k,r) - 1| > \frac{1}{2}$ for $|k| > \rho_r$. For such large $|k|$, the
argument of the logarithm in $i \beta_\ell(k,r) = \ln F_\ell(k,r)$
is entirely contained in the circle around unity of radius $1/2$,
where the logarithm is holomorphic. The absolute convergence of
$\sum_{\nu=1}^\infty f_\ell^{(\nu)}(k,r)$ thus implies the convergence of the
Born series for $\beta_\ell(k,r)$ outside a semi-circle of sufficiently large
radius $|k| > \rho_r$. Thus the (Born) series 
$\sum_{\nu=1}^m\beta_\ell^{(\nu)}(k,r) \equiv \beta_\ell(k,r)$ converges 
absolutely and uniformly at sufficiently large $|k|$ in the upper half plane,

Finally, we need to show that the difference $|\beta_\ell(k,0) -
\sum_{\nu=1}^m \beta_\ell^{(\nu)}(k,0)|$ vanishes at least as 
$\mathcal{O}(1/|k|^{2m+1})$ for large $|k|$ in the upper half $k$-plane.
We proceed inductively from eq.~(\ref{betaintell}) using integration by 
parts. What we will actually show is that the approximation 
$|\beta_\ell^{(\nu)}(k,r)|$ decays as $|k|^{-2\nu+1}$.  From the convergence 
of the Born series it is then clear that the remainder 
$|\ln F - \ln F_m| = |\beta_\ell - \sum_{\nu=1}^m\beta^{(\nu)}_\ell |
= |\sum_{m+1}^\infty \beta^{(\nu)}_\ell|$
vanishes at least as the leading term $\beta_\ell^{(m+1)}(k,r) 
\sim |k|^{-(2m+1)}$.
To estimate the large-$k$ behavior of $\beta^{(\nu)}_\ell(k,r)$, we rewrite 
the integral equation, eq.~(\ref{betaintell}) in the form
\begin{equation}
	\beta_\ell(k,r) = i \int_r^\infty dr_{1}\, \mathsf{K}_\ell(k,r_{1})
\end{equation}
where
\begin{equation}
	\mathsf{K}_\ell(k,r) \equiv \int_r^\infty
	dr_{1}\,\left(\frac{w_\ell(kr_{1})}	{w_\ell(kr)}\right)^2
	\Gamma_\ell(k,r_{1}) =	\int_r^\infty
	dr_{1}\,\exp\left[2ik\int_r^{r_{1}}\eta_\ell(kr_{2})\,
	dr_{2}\right] \Gamma_\ell(k,r_{1})\,.
\end{equation}
Integrating by parts once and estimating the remainder by the Riemann-Lebesgue
lemma, the leading asymptotic behavior of the kernel
$\mathsf{K}_\ell$ is easily found to be
\begin{equation}
	 \mathsf{K}_\ell(k,r) = 
        \Gamma_\ell(k,r)\left[\frac{1}{2 i k \eta_\ell(kr)}+ 
        \mathcal{O}(k^{-2})\right]\,.
\label{gap}
\end{equation}
From this estimate and the limit $\lim\limits_{|z|\to \infty} \eta_\ell(z) = 1$ 
in the upper half plane, we infer
\begin{equation}
	\beta^{(\nu)}_\ell(k,r) = \frac{1}{2 k}\int_r^\infty dr_{1}\,
	\Gamma^{(\nu)}_\ell(k,r_{1})
	\left[1 + \mathcal{O}(k^{-1})\right] \, .
\label{betaintagain}
\end{equation}

The starting iteration for the source is $\Gamma^{(1)}_\ell(k,r) = V(r)$,
which is independent of $k$ (and $\ell$), whence
\begin{equation}
\beta_\ell^{(1)}(k,r) = \frac{1}{2 k}\int_r^\infty dr_{1}\,
V(r_{1})\,\left[1+ \mathcal{O}(k^{-1})\right]\,.
\end{equation}
As a check, the large $k$ behavior of the first Born approximation 
in the $s$-channel is correctly predicted as 
\begin{equation}
	\delta_0^{(1)}(k) = - \,\mathsf{Re}\,\beta^{(1)}_0(k,0) = 
	-\frac{1}{2k} \int_0^\infty dr\,V(r) + \mathcal{O}(k^{-2})\,.
\label{gap2}
\end{equation}
For the next iteration, we note $\Gamma_\ell^{(2)}(k,r) = 
\left[\frac{d}{dr}\beta_\ell^{(1)}(k,r)\right]^2 \sim k^{-2}$
(with a real constant factor), so that
$ 
|\beta_\ell^{(2)}(k,r)|\sim |k|^{-3}\,.
$
The general iteration step follows from the assumption that 
$|\beta^{(\mu)}_\ell(k,r)| \sim |k|^{1- 2 \mu}$ for $\mu < \nu$.  
To find the behavior of $\beta^{(\nu)}_\ell(k,r)$, we need to 
consider the $\nu^{\rm th}$ order term $\Gamma_\ell^{(\nu)}(k,r)$
that is given in eq.~(\ref{Gammaexp}). By assumption, each 
$|\beta^{(\sigma)}_\ell(k,r)|$  decays as $|k|^{1-2\sigma}$ at large $k$.
Hence all the terms in eq.~(\ref{Gammaexp}) are of order 
$|k|^{1-2\sigma} |k|^{1-2\tau}=|k|^{2(1-\nu)}$. Thus the 
source $|\Gamma^{(\nu)}_\ell(k,r)|$ vanishes as $|k|^{2(1-\nu)}$. From
eq.~(\ref{betaintagain}), we easily complete the induction,
$|\beta_\ell^{(\nu)}(k,r)| \sim |k|^{1-2\nu}$, as claimed.

It should be noted that the $\mathcal{O}(k)$ estimates in eq.~(\ref{gap})
-- (\ref{gap2}) require the potential $V(r)$ to be bounded and sufficiently 
smooth to allow for integration by parts. Moreover, the Riemann-Lebesgue
lemma imposes certain restrictions on $V$ and its derivatives, as 
discussed in the main text. These restrictions are certainly satisfied 
for smooth bounded potentials from the Faddeev class (\ref{faddeev}), 
but more general cases such as a step function may also be handled.

Having established these two properties of the $\{\beta^{(\nu)}(k)\}$, 
it is clear that the auxiliary function satisfying (a) and (b) 
in the previous subsection takes the same form as in the 
antisymmetric channel in one dimension,
\begin{equation}
	F_{\ell,m}(k)= 
	\exp\left[i\sum_{\nu=1}^{m}\beta_{\ell}^{(\nu)}(k)\right].
	\label{auxell}
\end{equation}
The required properties of $F_{m}(k)$ follow directly from those of
$\beta_{\ell}^{(\nu)}(k)$ and the convergence of the Born series for 
sufficiently large $|k|$.  The quantity that enters the sum rule is 
$\delta_{\ell,m}(k)$, given by eq.~(\ref{deltalm}).  Clearly 
$\delta_{\ell,m}(k)$ is the sum of the first $m$ terms in the Born expansion 
of $\delta_{\ell}(k)$.  This completes the derivation of the sum rules in 
$\ell^{\rm th}$ partial wave in three dimensions.

\end{document}